\documentclass[prl,twocolumn,amsmath,amssymb,floatfix]{revtex4-1}
\usepackage{color,graphicx}
\usepackage{mathptmx}
\allowdisplaybreaks
\usepackage{hyperref}
\usepackage{multirow}
\usepackage{hhline}
\newcommand{\Lcal}{\mathcal{L} }

\begin{document}

\title{Improved effective linearization of nonlinear Schr\"odinger waves by increasing nonlinearity}
\author{Katelyn Plaisier Leisman$^{1}$, Douglas Zhou$^{2}$, J.~W.~Banks$^{3}$, Gregor Kova\v ci\v c$^{3}$, and David Cai$^{2,4}$}
\affiliation{\normalsize{$^{1}$ University of Illinois, Department of Mathematics, Urbana, IL 61801, USA}\\
\normalsize{$^{2}$ School of Mathematical Sciences, MOE-LSC, and Institute of Natural Sciences, Shanghai Jiao Tong University,  Shanghai 200240, People's Republic of China}\\
\normalsize{$^{3}$ Rensselaer Polytechnic Institute, Department of Mathematical Sciences, Troy, NY 12180, USA}\\
\normalsize{$^{4}$ Courant Institute of Mathematical Sciences, New York University, New York,  New York 10012, USA}}

\date{\small\today}

\begin{abstract}
From among the waves whose dynamics are governed by the nonlinear Schr\"odinger (NLS) equation, we find a robust, spatiotemporally disordered family, in which waves initialized with increasing amplitudes, on average, over long time scales, effectively evolve as ever more weakly coupled collections of plane waves.   In particular, the relative amount of energy contained in their coupling decays to zero with increasing wave amplitude.
\end{abstract}

\maketitle

Linear autonomous conservative wave systems are characterized by the absence of higher-harmonic generation, by the linearity of their governing equations, and by the quadratic total energy.   Nonlinear wave systems, in turn, are characterized by pronounced generation of higher harmonics, frequently via nonlinear coupling in the governing equations, and by the total energy with a higher-than-quadratic nonlinearity at the leading order reflecting this coupling.  Typically, the nonlinearity attributes are expected to increase with the amplitude of the physical variable determining the dynamics of a nonlinear system.  (See Ref.~\cite{satsuma74} for a striking example.)
In this letter, we illustrate on the example of the NLS equation~\cite{benneynewell67,zakharov68,zakharov72,zakharov72,hasegawa73},
\begin{equation}
  i \psi_t =\psi_{xx}\pm 2|\psi|^2 \psi ,\label{eq:nls}
\end{equation}
that such an increase does not necessarily occur. [Hereafter, the upper sign corresponds to the \emph{focusing} NLS (FNLS) and lower to the \emph{defocusing} NLS (DNLS).] Instead, in a statistical sense, a robust family of spatiotemporally disordered NLS waves  exhibits properties that remain close to linear, and, in fact, some behavior becomes increasingly linear with increased wave amplitude.
  Specifically, we show that such waves behave, on average, over long times, as ever more weakly coupled collections of plane waves, and that the energy contribution of this coupling decreases to zero with increasing wave amplitude.

Equation~(\ref{eq:nls}) is a completely integrable~\cite{zakharov72}, universal envelope equation~\cite{benneynewell67},  valid for describing weakly-nonlinear, modulated, single-frequency wavetrains in phenomena such as surface waves~\cite{zakharov68} and nonlinear optics~\cite{hasegawa73}.   We note that, consequently, NLS waves with exceedingly high amplitudes have not been observed or measured experimentally.   Instead, to find a physical realization of the effect described here, we can take advantage of the scaling symmetry 
\begin{equation} x\to \lambda x,\quad t\to \lambda^2 t, \quad \psi \to \lambda\psi, \quad \lambda>0. \label{eq:symm}
\end{equation} 
This symmetry leaves the NLS equation in Eq.~(\ref{eq:nls}) intact, and implies that increasing the wave amplitude and fixing the spatial extent of the wave is equivalent to fixing the wave amplitude and increasing the spatial extent of the wave.   In other words, we arrive at the following equivalently paradoxical statement: the farther a particular disordered NLS wave with a fixed amplitude propagates, the more it  becomes a collection of linearly evolving plane waves in a statistical sense.    Thus, the effect we describe could perhaps be observed using data obtained with techniques such as those developed for studying disordered and ``rogue" wave envelopes in long wave tanks and nonlinear optics~\cite{onorato05,ONORATO2006586,ONORATO201347,suret16,tikan2021nonlinear}.  For example, one could try to observe it using a sequence of periodically-repeating, disordered optical signals of increasing duration, as further described in the penultimate paragraph of this letter.

For a disordered wave belonging to the family referred to in the previous two paragraphs, on average, over long time scales, the dynamics of any individual mode in the wavenumber space are known to be close to sinusoidal, oscillating with an \emph{effective frequency} that depends on the size of the wave~\cite{erdogan08,lee09,Leisman19a,tikan2021nonlinear}.   
This effective frequency can be used to decompose the NLS equation into effective linear and effective nonlinear parts, or similarly to decompose the NLS Hamiltonian into effective quadratic and effective superquadratic parts. In this Letter, we show that this decomposition minimizes the sizes of the respective effective nonlinear components of the NLS equation and the superquadratic components of the Hamiltonian, as the underlying disordered NLS waves evolve over long times. We also show that the ratio of the effective nonlinear and linear parts of the NLS equation approaches a finite limit, including possibly a vanishing limit, in the limit of large amplitudes. Furthermore, and most importantly, we show that the ratio of the effective superquadratic and quadratic parts of the NLS Hamiltonian decays to zero. 
Therefore, for such waves, as we increase their amplitudes, on average, over long times, their single-mode, plane-wave components dominate, while the nonlinear, mode-coupling components remain bounded or even vanish in comparison.   In fact,  with increasing wave amplitudes, the energy content of the mode-coupling components always vanishes in comparison with the energy content of the plane-wave components for the waves belonging to this family.   Analogous results also hold for the equivalent case of increasing spatial extent of the waves.

\begin{table*}[htbp]
\begin{tabular}{ccccccccccc}\cline{1-11}
\multicolumn{10}{ l }{\bf IC1} \\
\multicolumn{1}{ c  }{}                        &
\multicolumn{1}{ p{3.5cm}}{{Amplitude
 $A$}} & 3.2 &  6.4 & 12.7 & 25.5 & 50.9 & 101.8 & 203.6 & 407.3 & 814.6     \\ \cline{1-11}
\multicolumn{1}{ c  }{}                        &
\multicolumn{1}{ l }{Norm $\|\psi\|$} & 8.0 & 16.0 & 31.9 & 63.8 & 127.6 & 255.2 & 510.5 & 1020.9 & 2041.9     \\ \cline{1-11}
\multicolumn{1}{ c  }{}                        &
\multicolumn{1}{ l }{End Time} & 128$\pi$ & 32$\pi$ & 8$\pi$ & 2$\pi$ & $\pi$/2 & $\pi$/8 & $\pi$/32 & $\pi$/128 & $\pi$/512   \\ \hhline{===========}
\multicolumn{10}{ l  }{\bf IC2} \\
\multicolumn{1}{ c  }{}                        &
\multicolumn{1}{ l }{Amplitude $A$} & 1.5 &  2.1 &  3.0 &  4.2 &  6.0 &  8.5 & 12.0 & 17.0 & 24.0 \\ \cline{1-11}
\multicolumn{1}{ c  }{}                        &
\multicolumn{1}{ l }{Norm $\|\psi\|$} & 4.2 &  5.6 &  7.6 & 11.4 & 15.3 & 21.4 & 30.3 & 42.6 & 60.3 \\ \cline{1-11}
\multicolumn{1}{ c  }{}                        &
%\multicolumn{1}{ l }{End Times for Fig. 1} & 8$\pi$ & 4$\pi$ & $\pi$ & $\pi$/2 & $\pi$/8 & $\pi$/16 & $\pi$/64 & $\pi$/128 & $\pi$/512 \\ \cline{1-11}
%\multicolumn{1}{ c  }{}                        &
\multicolumn{1}{ l }{End Times} &  128$\pi$ & 64$\pi$ & $32\pi$ & $16\pi$ & $8\pi$ & $4\pi$ & $2\pi$ & $\pi$ & $\pi$/2 \\ \cline{1-11}
\end{tabular}
\caption{Parameter values used to generate the waves presented in the figures.  Note that the end times are inversely proportional to the square of the amplitude $A$ as mentioned in the text.  In particular, for IC1, the end times are $\sim 4000/A^2$, and for IC2, they are $\sim 900/A^2$.}\label{tbl:norms}
\end{table*}

In the case of increasing amplitude, the above robust family of disordered NLS waves is noise-like in that the waves it contains have increasingly short-range spatial correlations, and also short-range correlations in wave number space~\cite{yaglom}.
In this Letter, we restrict our consideration to the 
spatially periodic case with fixed period $L$.  (In all the figures, we use $L=2\pi$.)  For FNLS, the initial condition for disordered waves can be prepared in two different ways:   \textbf{IC1}: We exploit the modulational instability of FNLS traveling or standing waves, $A e^{i [\gamma x + (\gamma^2-2A^2)t]}$, with $\gamma$ an integer multiple of $2\pi/L$, to small perturbations~\cite{hasimoto72,weideman86}.  We assume the initial wave in the form $Ae^{i\gamma x}\left(1+\sum_{-N}^{N} \varepsilon_ne^{2\pi i n x/L}\right)$, where $N$ is the largest integer such that $1\le N\le AL/\pi$ and $\varepsilon_n$ are complex amplitudes with magnitudes $|\varepsilon_n|$ uniformly distributed in the interval $[0 ,\varepsilon]$, for some small $\varepsilon>0$, and phases uniformly distributed on $[0,2\pi]$. (We use $\varepsilon = 10^{-3}\sqrt{3/2A\sqrt{2}}$ and $\gamma=0$, 10, and 100.)  
\textbf{IC2}: We prepare random initial wave-forms $\sum_{-M}^{M} c_ne^{2\pi i n x/L}$, where $M$ is the largest integer $\leq \sqrt{2}A^2L/\pi$, and the real and imaginary parts of the coefficients $c_n$ 
are drawn from the Gaussian distribution with zero mean and variance $A^2/[2(2M+1)]$.  These forms render a smooth approximation to spatial white noise with the minimal wavelength $\pi/A^2$ and spatial standard deviation $A$ for increasing $A$~\cite{filip19}.   For DNLS, we use \textbf{IC2}.   In what is to follow, we will loosely refer to the parameter $A$ as the amplitude of the wave $\psi(x,t)$ generated from either IC1 or IC2.
\begin{figure}[htbp]
\centering\includegraphics[trim=0cm 0cm 0cm 0cm, clip=true,width=.48\textwidth]{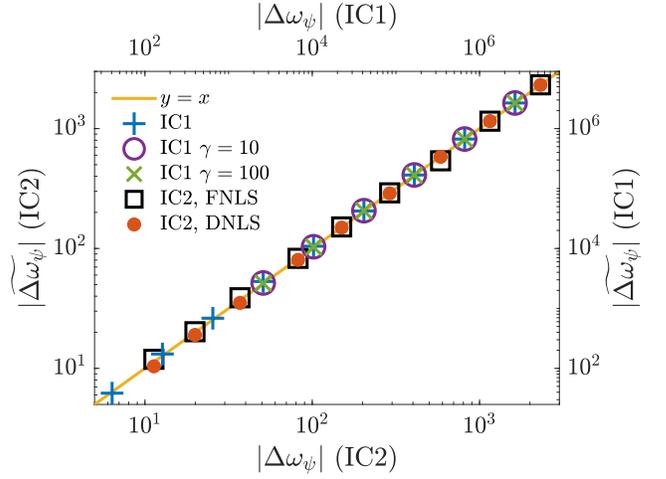}
\caption{Scatter plot of the frequency shift $\Delta\omega_\psi$, predicted by the EDR in Eq.~(\ref{eq:edr}), versus the corresponding frequency shift $\widetilde{\Delta\omega}_\psi$, obtained using the WFS method, for increasing values of the wave norm. }\label{fig:scatter}
\end{figure}

%\begin{figure*}[htbp]
%\centering\includegraphics[trim=0cm 0cm 0.72cm 0cm, clip=true,width=.32\textwidth]{figures/LeismanFig2a3}
%\centering\includegraphics[trim=0cm 0cm 0.72cm 0cm, clip=true,width=.32\textwidth]{figures/LeismanFig2b}
%\centering\includegraphics[trim=0cm 0cm 0.72cm 0cm, clip=true,width=.32\textwidth]{figures/LeismanFig2c}
%\caption{
%Time-averaged norm of the trial renormalized nonlinear term in the NLS as a function of the renormalization parameter, $\beta$,  for increasing norms of the NLS waves.  (a) FNLS waves generated using IC1; (b) FNLS waves generated using IC2.  (c) DNLS waves generated using IC2.  \red{We use standing waves, $\gamma=0$, for IC1; the results for traveling waves with wavenumbers $\gamma=10$ and 100 appear almost identical.}}\label{fig:beta}
%\end{figure*}
\begin{figure}[htbp]
\centering\includegraphics[trim=0cm 0cm 0cm 0cm, clip=true,width=.48\textwidth]{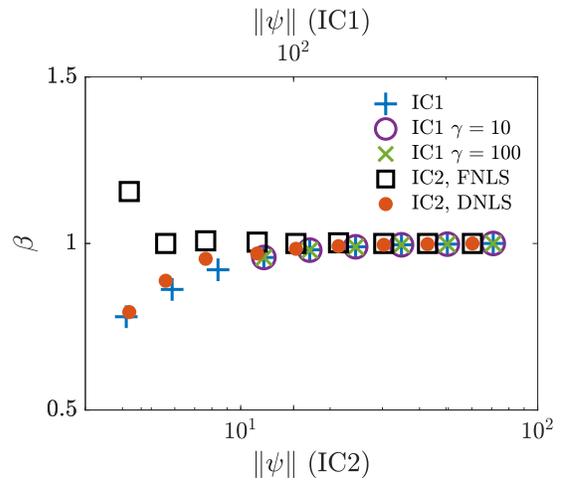}
\caption{
The values of the renormalization parameter, $\beta$, that minimize the time-averaged norm of the trial renormalized nonlinear term, $\mathcal{N}_\beta$, in the NLS  for increasing norms of the NLS waves. }\label{fig:beta}
\end{figure}
\begin{figure}[htbp]
\centering\includegraphics[trim=0cm 0cm 0cm 0cm, clip=true,width=.48\textwidth]{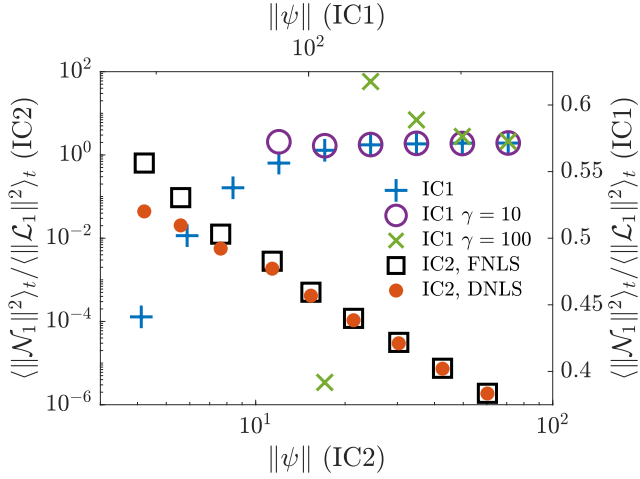}
\caption{
Ratio between the time-averaged norms of the effective nonlinear and linear NLS terms for waves with increasing norms.  The values of the wave norms are listed in Table~\ref{tbl:norms}.   
}\label{fig:rhsratio}
\end{figure}

As mentioned above, one linear-like property of disordered NLS waves is an \emph{effective dispersion relation} (EDR), derived in the small-amplitude and long-wavelength limits~\cite{erdogan08,lee09,Leisman19a}, and confirmed for all amplitude and wavelength sizes using the wavenumber-frequency-spectral (WFS) method by finding a sharp peak at a unique effective frequency $\omega_k$ in the power-spectral density of the evolving mode of the wave $\psi(x,t)$ with wave number $k$~\cite{wakata07,farrar08,Shinoda09,shinoda10}.   This EDR for NLS waves was also measured and analyzed experimentally in water waves~\cite{tikan2021nonlinear}.
Specifically, for each wave number $k$, the effective frequency is given by the equation
\begin{equation}\omega_k=k^2\mp \frac{4}{L}\Vert \psi\Vert^2, \label{eq:edr}\end{equation} 
where $\Vert\psi\Vert^2 = \int_{-L/2}^{L/2} |\psi(x,t)|^2\,dx$ is the (time-conserved) squared norm of the wave $\psi(x,t)$. 
Note that $\Vert\psi\Vert/\sqrt{L}$ is the average amplitude of the wave, and that it is $\approx A$ for IC1 and approaches $A$ as it increases for IC2.

We should remark that $k^2$ in Eq.~(\ref{eq:edr}) is the frequency obtained from the corresponding linear dispersion relation of the Schr\"odinger equation for a free particle.  In other words, under the influence of the quadratic self-interaction potential, a noisy NLS wave effectively evolves as it would under the linear Schr\"odinger dynamics for a free particle, but with an additional, linearly increasing temporal phase due to the wavenumber-independent effective frequency shift $\Delta\omega_\psi=\mp 4\Vert \psi\Vert^2/L $, obtained from Eq.~(\ref{eq:edr}).   The results in this Letter give strong evidence that this picture improves, statistically, with increasing wave amplitude.

To obtain the results presented in this Letter, we have performed well-resolved numerical simulations of the governing NLS equation in Eq.~(\ref{eq:nls}). Our approach uses fully coupled, 6th-order accurate Runge-Kutta time stepping with spectrally accurate spatial derivatives computed via  Fast Fourier Transform~\cite{fornberg78}.
 We have used great care to ensure that the simulations are fully resolved, and thus the results are insensitive to numerical considerations. In particular, we have compared our results to those from similarly resolved calculations using a 4th-order accurate Runge-Kutta scheme, as well as to more coarsely resolved simulations from the 4th and 6th order schemes, and the results we present are essentially unchanged. Furthermore, the simulation results presented here preserve the first three NLS conserved quantities~\cite{zakharov72} to five digits of relative precision.  (The first two of these quantities are the wave norm $\Vert \psi\Vert$ and the Hamiltonian $H$ in Eq.~(\ref{eq:ham}) below.)   
 
\begin{figure}[htbp]
\centering\includegraphics[trim=0cm 0cm 0cm 0cm, clip=true,width=.48\textwidth]{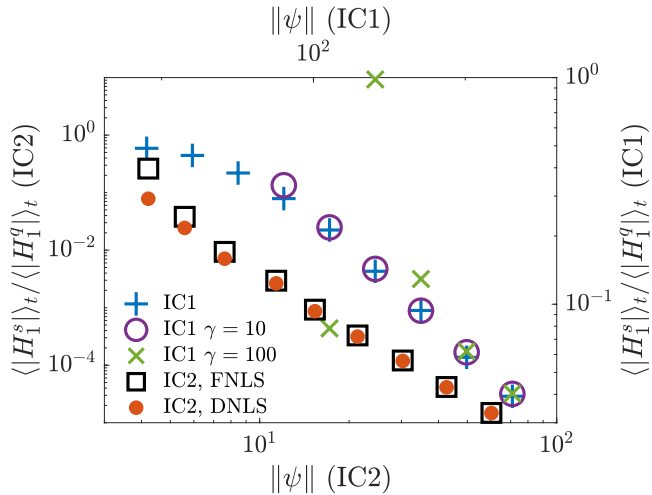}
\caption{
Ratio between the time-averaged sizes of the effective superquadratic and quadratic parts of the NLS Hamiltonian for waves with increasing norms.  The values of the wave norms are listed in Table~\ref{tbl:norms}.    
}
\label{fig:hamratio}
\end{figure}

The amplitudes $A$, norms $\Vert\psi\Vert$ of the waves, and end times of the time intervals over which we evolved the waves used in producing all the figures are displayed in Table \ref{tbl:norms}. Note that the end times we use scale as $O(1/A^2)$ with the wave amplitude, $A$, as suggested by the scaling in Eq.~\eqref{eq:symm}.  Our results, in particular those presented in Figs.~\ref{fig:rhsratio} and~\ref{fig:hamratio}, have converged to their respective limits by these listed end times.

In Fig.~\ref{fig:scatter}, for three families of disordered NLS waves with increasing amplitudes, we confirm excellent agreement in the scatter plots of the frequency shifts $\Delta\omega_\psi$ 
obtained from Eq.~(\ref{eq:edr}) versus the corresponding shifts $\widetilde{\Delta\omega}_\psi $ obtained  using WFS analysis.

%\begin{figure*}[htbp]
%\centering\includegraphics[trim=0cm 0cm 0.72cm 0cm, clip=true,width=.32\textwidth]{figures/LeismanFig4a3}
%\centering\includegraphics[trim=0cm 0cm 0.72cm 0cm, clip=true,width=.32\textwidth]{figures/LeismanFig4b}
%\centering\includegraphics[trim=0cm 0cm 0.72cm 0cm, clip=true,width=.32\textwidth]{figures/LeismanFig4c}
%\caption{
%Time-averaged size of the trial renormalized superquadratic part of the NLS Hamiltonian as a function of the renormalization parameter, $\beta$,  for increasing norms of the NLS waves.  (a) FNLS waves generated using IC1; (b) FNLS waves generated using IC2.  (c) DNLS waves generated using IC2.  \red{We use standing waves, $\gamma=0$, for IC1; the results for traveling waves with wavenumbers $\gamma=10$ and 100 appear almost identical.}}
%\label{fig:betaham}
%\end{figure*}
\begin{figure}[htbp]
\centering\includegraphics[trim=0cm 0cm 0cm 0cm, clip=true,width=.48\textwidth]{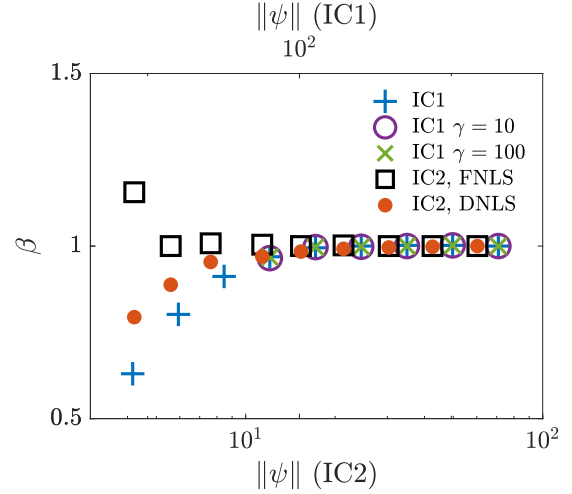}
\caption{
The values of the renormalization parameter, $\beta$, that minimize the time-averaged size of the trial renormalized superquadratic part of the NLS Hamiltonian, $H_\beta^s$,    for increasing norms of the NLS waves.  }
\label{fig:betaham}
\end{figure}
In the next two paragraphs we present the main results of this letter.  
We begin by verifying that the choice of the frequency given by the EDR in Eq.~(\ref{eq:edr}) minimizes the effective nonlinear portion of the NLS wave evolution equation.
To carry out this minimization, for each mode with wavenumber $k$, we assume a trial \emph{renormalized frequency}, $k^2\mp 4\beta \Vert\psi\Vert^2/L$, obtained via renormalizing the linear frequency $k^2$ by adding to it the product of the frequency shift $\Delta\omega_\psi$ from  Eq.~(\ref{eq:edr}) with the minimization parameter $\beta$.  Equivalently, we rewrite the right-hand side of the NLS in Eq.~(\ref{eq:nls})
as 
\begin{align}i\psi_t&=\left(\psi_{xx}\pm \frac{4\beta}{L} \Vert\psi\Vert^2\psi \right)\pm \left(2|\psi|^2 \psi -\frac{4\beta}{L} \Vert\psi\Vert^2\psi\right) \notag\\&\equiv \Lcal_\beta + \mathcal{N}_\beta, \label{eq:splitfield}\end{align}
where $\Lcal_\beta$ and $\mathcal{N}_\beta$ are the trial  \emph{renormalized linear} and \emph{nonlinear} terms.  
 In Fig.~\ref{fig:beta}, for increasing values of the wave norm $\Vert\psi\Vert$, we display the $\beta$-values that minimize the time-averaged size of the trial renormalized nonlinear term, $\left\langle\Vert \mathcal{N}_\beta \Vert^2\right\rangle_t$,  which represents the size of the (formal) nonlinearity.   We see that these $\beta$-values approach $\beta = 1$, for which the trial renormalized frequency becomes the true effective frequency, $\omega_k$ in Eq.~(\ref{eq:edr}).  In Fig.~\ref{fig:rhsratio} we display the dependence of the ratio $\left\langle\Vert \mathcal{N}_1 \Vert^2\right\rangle_t/\left\langle\Vert\Lcal_1 \Vert^2\right\rangle_t$ between the time-averaged norms of the \emph{effective nonlinear} and \emph{linear} terms in the evolution of the NLS wave as a function of the wave norm $\Vert\psi\Vert$. 
For FNLS waves prepared using IC1, we see that this ratio saturates near the value 0.57 for large $\Vert\psi\Vert$, and does so for all three values of the wavenumber $\gamma$ used in our simulations.  It therefore does not grow indefinitely with the size of the nonlinearity. For both FNLS and DNLS waves prepared using IC2, in turn, this ratio decays to zero,  so that the effective linear term dominates in the large-amplitude limit.

Finally, we consider the Hamiltonian~\cite{zakharov74} 
\begin{equation}H= \Vert \psi_x\Vert^2\mp \Vert\psi^2\Vert^2,\label{eq:ham}\end{equation}
from which the NLS in Eq.~(\ref{eq:nls}) is derived via the formula $i\psi_t = \delta H/\delta \psi^*$, where $\delta$ denotes the variational derivative and $^*$ denotes complex conjugation.  We split this Hamiltonian into its trial \emph{renormalized quadratic} and \emph{superquadratic} parts, 
\begin{align}H&= \left(\Vert \psi_x\Vert^2\mp \frac{2\beta}{L} \Vert \psi\Vert^4\right)\mp \left(\Vert\psi^2\Vert^2 -\frac{2\beta}{L} \Vert \psi\Vert^4\right)\notag\\&\equiv H_\beta^q+H_\beta^s,\label{eq:hamil}\end{align}
where the trial renormalized quadratic part $H_\beta^q$ of the Hamiltonian $H$ generates the renormalized linear term $\Lcal_\beta$ in the evolution of the wave $\psi$ in Eq.~(\ref{eq:splitfield}) and the trial renormalized superquadratic part $H_\beta^s$ of the Hamiltonian $H$ generates the trial renormalized nonlinear term $\mathcal{N}_\beta$.   
In Fig.~\ref{fig:betaham}, for increasing values of the wave norm $\Vert\psi\Vert$, we display the $\beta$-values for which the time-averaged trial renormalized superquadratic part of the Hamiltonian, $\left\langle \left\vert H_\beta^s\right\vert\right\rangle_t$, is minimized.  We again find that this part is minimal as $\beta$ approaches $\beta=1$, which is when the mode with wavenumber $k$  oscillates with the effective linear frequency $\omega_k$ in Eq.~(\ref{eq:edr}). In Fig.~\ref{fig:hamratio} we then display the dependence of the time-averaged ratio $\left\langle \left\vert H_1^s \right\vert \right\rangle_t/\left\langle \left\vert H_1^q\right\vert \right\rangle_t$ between the sizes of the \emph{effective superquadratic} and \emph{quadratic} parts of the Hamiltonian $H$ in Eq.~(\ref{eq:hamil}) as a function of the wave norm $\Vert\psi\Vert$.
We see that this ratio decays to zero for large $\Vert\psi\Vert$ for all the cases we considered.   In other words, for large nonlinearities, the effective superquadratic part of the Hamiltonian becomes negligible as compared to its effective quadratic part.   

The discussion in the above two paragraphs confirms that a highly nonlinear and disordered NLS wave effectively evolves as a collection of weakly-coupled plane waves governed by the EDR in Eq.~(\ref{eq:edr}), as discussed in Refs.~\cite{erdogan08,lee09,Leisman19a}, but now for a range of NLS wave amplitudes.   We reemphasize that the main result of this Letter consists of demonstrating that the relative energy contained in this coupling decreases to zero with increasing nonlinearity.    Moreover, for waves prepared using IC2, the strength of this coupling decreases to zero even when measured by the relative size of the effective nonlinear versus linear terms in the NLS equation itself.   We should reiterate that the same results, with obvious modifications, also hold in the case when the wave amplitude $A$ is held fixed and the spatial period $L$ is increased instead.

Curiously, for an ensemble of initial conditions IC2 with a given amplitude $A$, we can explicitly estimate the sizes of the effective linear and nonlinear terms,  $\Vert\Lcal_1 \Vert^2$ and $\Vert \mathcal{N}_1 \Vert^2$ in Eq.~(\ref{eq:splitfield}) with $\beta=1$,  using properties of the Gaussian distribution.   We find $\Vert\Lcal_1 \Vert^2\sim O(A^{10})$ and $\Vert \mathcal{N}_1 \Vert^2\sim O(A^6)$.    Their ratio scales as $\sim O(A^{-4})$, which agrees with the asymptotic slope of the data in Fig.~\ref{fig:rhsratio}.  Likewise, for the effective quadratic and superquadratic terms, $\left\vert H_1^q \right\vert$ and  $\left\vert H_1^s \right\vert$ in the Hamiltonian in Eq.~(\ref{eq:hamil}) with $\beta=1$, we find the estimates $\left\vert H_1^q \right\vert \sim O(A^6)$ and  $\left\vert H_1^s \right\vert \sim O(A^3)$, and so their ratio scales as $\sim O(A^{-3})$, which agrees with the asymptotic slope of the data in Fig.~\ref{fig:hamratio}.    We remark that the scaling of the superquadratic term $\left\vert H_1^s \right\vert$ crucially depends on the form of the EDR in Eq.~(\ref{eq:edr}), which induces a nontrivial cancellation of terms scaling as $\sim O(A^4)$.  The not-at-all-obvious fact that these scalings of the initial conditions effectively persist when the resulting waves evolve under the NLS dynamics suggests that, statistically, the form of the evolving wave indeed remains the same as IC2, just with the phases of the individual plane waves evolving linearly according to the EDR in Eq.~(\ref{eq:edr}).   In other words, our simulations suggest that a wave emerging from an initial wave of the form IC2 statistically samples other realizations of IC2 with the same parameters.

The concept of the EDR stems from the earlier concept of the nonlinear frequency shift~\cite{zakharov92}.    It has been well studied both experimentally and theoretically~\cite{Ohe_1981b,herbers02,gershgorin05,gershgorin07,PhysRevLett.105.144502,lee13,shixiao14,Manktelow:2014hc,shiziao_16,Jiang:2016dz}, including for the NLS equation~\cite{erdogan08,lee09,Leisman19a}, and is important for the understanding of wave-wave interactions and energy transfer in turbulent dispersive-wave  systems.  Methods for deriving EDRs analytically are described in these works, and the procedure most frequently used to compute it numerically is the WFS method~\cite{wakata07,farrar08,Shinoda09,shinoda10}.   In general, one should expect an EDR to only hold approximately, and increasingly less so with the increasing wave amplitude.   We reiterate that, in this Letter, in turn, we have presented a particularly striking, counterintuitive example, in which the EDR not only holds but even improves and becomes ``exact" in the limit of large amplitudes of certain NLS waves in the manner described above.

We should not, however, be mislead into believing that high-amplitude solutions could transform the NLS into a genuinely linear system.  The crucial missing property is linear superposition.   Namely, unless two high-amplitude NLS waves, as well as all their linear combinations, share the same norm, which is impossible, they do not even share the same EDR and so the two waves and each of their linear superpositions approximately satisfy three different effective linear equations.  The nonlinearity of the NLS waves is also clearly reflected in the dependence of the effective frequency, $\omega_k$ in Eq.~(\ref{eq:edr}), on the wave norm $\Vert\psi\Vert$.  Nevertheless, as opposed to a repeatable experiment in which initial waves can be created at will, an observation of a single wave cannot be used to verify linear superposition.   In fact, any observed wave evolving as a sum of (effectively) decoupled plane waves without generating harmonics appears linear.   For example, from the WFS method~\cite{wakata07,farrar08,Shinoda09,shinoda10} alone, one cannot discern whether the system that gave rise to the observed waves was linear or nonlinear.    Our results thus lead to the paradoxical conclusion that, at large amplitudes, two effectively linear NLS waves may add nonlinearly to a third effectively linear NLS wave.

We now briefly discuss whether the linearization phenomenon described here could, in principle, be measured in an optical system.    For example, let us consider a standard monomode optical fiber in the transparency window for the wavelength $\lambda = 1.55~\mu$m.  Since the roles of $x$ and $t$ are reversed for the NLS in fiber optics~\cite{agrawal13}, a potential experiment  would inject a time-periodic signal containing quasi-random noise in the sense of IC1 or IC2 into one end of the fiber,  and read the propagated signal at the opposite end.   To keep the signal intensity bounded, one would use the symmetry in Eq.~\eqref{eq:symm} and study the wave envelope $\psi(x/A,t/A^2)/A$ injected as the appropriately scaled IC1 or IC2, where $A$ is the amplitude parameter in Table~\ref{tbl:norms} and $\psi$ the wave corresponding to that amplitude.  Ensemble averaging would replace the time averaging employed in this letter.   Using the parameters listed in Ref.~\cite{agrawal13}, we conclude that, for IC1, injected signals with characteristic power $\approx 0.3$ Watt and temporal periods $\approx 15\times A $ picoseconds would cover all the corresponding values of the parameter $A$ listed in Table~\ref{tbl:norms}.   Our simulation time of $\approx 4000/A^2$ would translate to a required propagation length of $\approx 8000$ kilometers.   Likewise, for IC2, the same characteristic power and the temporal periods $\approx 30 \times A $ picoseconds would again cover all the corresponding values of $A$ in Table~\ref{tbl:norms}.    Moreover, in this case, the smallest pulsewidth involved, $\approx 30/A$ picoseconds, would still exceed $\approx 1$ picosecond, so that the NLS description would remain valid even for the largest value of $A$ we have considered.
Our simulation time of $\approx 900/A^2$ would translate to a required propagation length of  $\approx 7200$ kilometers.    Both propagation lengths are comparable to those of transoceanic optical communication lines.   The main limiting obstacle would be losses, which are about 2.5\,\% per kilometer.   While the dispersion-loss compensation technique used in transoceanic lines~\cite{Gabitov:96,Gabitov:1996vc,Ablowitz:98} may distort the signal too much, distributed amplification, such as using fibers doped with the rare earth erbium~\cite{agrawal13}, or higher nonlinearity of the fiber material~\cite{Newhouse:90,127214,1159362,2006,2008} might help.   Far in the future, an experiment might perhaps take advantage of highly nonlinear exotic materials~\cite{SAHA2020} or metamaterials composed of transparent dielectrics and nano-sized metallic inclusions~\cite{Drachev:2018uz}.    Moreover, the factors of $\approx 4000$ and $\approx 900$ in our computational times were chosen arbitrarily, in order to facilitate time averaging.   Since an experiment wold have to use ensemble averaging instead, these factors might be reduced, resulting in more manageable propagation lengths.   
Nevertheless, the losses appear to present a technical rather than fundamental obstacle to the measurability of the improved linearization phenomenon discussed in this letter.  
 It is, however, important to note that our required computational times of $O(1/A^2)$ translate to $O(1)$ propagation distances for the signal regardless of the prefactor.   Longer computational times might lead to the effect not being measurable in principle.

Theoretically, the statistical linearization of NLS waves via mode decoupling is not all that surprising in light of the fact that the NLS equation in Eq.~(\ref{eq:nls}) is completely integrable~\cite{zakharov72}, and that its waves can therefore be described in terms of action-angle variables~\cite{zakharov74,fadeev87}.   What is surprising is that, in the large-amplitude limit, this linearization is achieved in terms of the \emph{linear} spatial modes.  The effective vanishing of the nonlinear coupling terms in the NLS equation and/or Hamiltonian at high wave amplitudes possibly follows from the absence of nontrivial resonances among
the effective frequencies of the NLS wave modes and the corresponding absence of nontrivial resonant wave-wave interactions, which is known to be connected to the complete integrability of the NLS equation~\cite{zakharov-schulman91}.  Therefore, we believe the improvement of the effective linearization with increasing wave amplitude that we described above may be quite rare.  Verifying this statement, for example on the general Majda-McLaughlin-Tabak model~\cite{majda97}, of which the two types of the NLS equation are  special cases, will be an interesting future project.

We are grateful to S. Jiang, A. Korotkevich, P. Kramer, K. Newhall, C. Schober, M. Schwartz, and V. Zharnitsky for helpful discussions.  This work was partly supported by US Dept.~of Education,  Student Innovation Center at Shanghai Jiao Tong University,
US NSF (Grants No. 0636358, 1344962, 1615859), US PECASE Award, NSFC (Grants No. 11671259, 11722107, 91630208, and 31571071), Simons Foundation, and SJTU-UM Collaborative Research Program.

\bibliography{edrbibfile}

\end{document}